\documentstyle[12pt]{article}
\title{Gaugino pair production at LHC (CMS)}
\author{\large S.I.~Bityukov~$^1$, N.V.~Krasnikov  \\[3mm]
\em Institute for Nuclear Research RAS, \\
\em Moscow, 117312, Russia  
}
\date{}
\begin{document}
\maketitle
\begin{abstract}
We investigate $\tilde \chi^{\pm}_1 \tilde \chi^0_2$ pair production
at LHC (CMS) with subsequent decays into leptons for the case 
of nonuniversal gaugino masses. Visibility of signal by an excess
over SM background in $3l + no~jets + E^{miss}_T$ events
depends rather strongly on the relation between LSP mass 
$\tilde \chi^0_1$ and $\tilde \chi^{\pm}_1$ mass.
\end{abstract}

\section{Introduction}

One of the LHC goals is the discovery of the supersymmetry. 
In particular, it is very important to investigate a possibility
to discover nonstrongly interacting superparticles (sleptons,
higgsino, gaugino). In ref.\cite{1}
 (see, also references \cite{2,3}) the LHC gaugino discovery potential
has been investigated within the minimal SUGRA-MSSM framework where all 
sparticle masses are determined mainly by two parameters: $m_0$ (common
squark and slepton mass at GUT scale) and $m_{1 \over 2}$ (common
gaugino mass at GUT scale). 
The signature used for the search for gauginos at LHC is 
$3~isolated~leptons + no~jets + E^{miss}_T$ events. 
The conclusion of these studies is that LHC is able 
to detect gauginos with $m_{1 \over 2}$ up to 150~GeV and in some
cases (small $m_0$) up to 400~GeV.

In this paper we investigate the gaugino discovery potential of LHC for the
case of nonuniversal gaugino masses. Despite
the simplicity of the  SUGRA-MSSM framework it is a very particular
model. The mass formulae for sparticles in  SUGRA-MSSM model are derived
under the assumption that at GUT scale ($M_{GUT} \approx 2 \cdot 10^{16}$~GeV) 
soft supersymmetry breaking terms are universal. However, in general,
we can expect that real sparticle masses can differ in a drastic way 
from sparticle masses pattern of SUGRA-MSSM model due to many reasons,
see for instance refs.~\cite{4,5,6,7}

\vspace{1cm}
\bigskip

\noindent
\rule{3cm}{0.5pt}\\
$^1$~~Institute for High Energy Physics, Protvino, Russia

\newpage

Therefore, it is more appropriate to investigate the LHC SUSY discovery 
potential in a model-independent way. The cross section 
for the $\tilde \chi^{\pm}_1 \tilde \chi^0_2$
chargino second neutralino pair production depends mainly on the mass of 
chargino which is approximately degenerate in mass with the second
neutralino $M(\tilde \chi^{\pm}_1) \approx M(\tilde \chi^0_2)$. 
The two lightest neutralino and the lightest chargino 
$(\tilde \chi^0_1, \tilde \chi^0_2, \tilde \chi^{\pm}_1)$ have, as largest
mixing components, the gauginos, and hence their masses are determined by
the common gaugino mass, $m_{1 \over 2}$. Within mSUGRA model
$M(\tilde \chi^0_1) \approx 0.4 m_{1 \over 2}$ and
$M(\tilde \chi^0_2) \approx M(\tilde \chi^{\pm}_1) 
\approx 2 M(\tilde \chi^0_1)$. 

The lightest chargino $\tilde \chi^{\pm}_1$ has several leptonic decay modes
giving an isolated lepton and missing energy:

three-body decay

\begin{itemize}

\item 
$\tilde \chi^{\pm}_1 \longrightarrow \tilde \chi^0_1  + l^{\pm} + \nu$,

\end{itemize}

two-body decays

\begin{itemize}

\item
$\tilde \chi^{\pm}_1  \longrightarrow  \tilde l^{\pm}_{L,R} + \nu$,

\hspace{16mm}  $\hookrightarrow \tilde \chi^0_1 + l^{\pm}$

\item
$\tilde \chi^{\pm}_1 \longrightarrow \tilde \nu_L + l^{\pm}$,

\hspace{16mm} $ \hookrightarrow \tilde \chi^0_1 + \nu$

\item
$\tilde \chi^{\pm}_1 \longrightarrow \tilde \chi^0_1 + W^{\pm}$.

\hspace{26mm} $ \hookrightarrow l^{\pm} + \nu$

\end{itemize}

Leptonic decays of $\tilde \chi^0_2$ give two isolated leptons and missing 
energy:

three-body decays

\begin{itemize}
\item 
$\tilde \chi^0_2 \longrightarrow \tilde \chi^0_1 + l^+ l^-$,

\item
$\tilde \chi^0_2 \longrightarrow \tilde \chi^{\pm}_1 + l^{\mp} + \nu$,

\hspace{16mm} $ \hookrightarrow \tilde \chi^0_1 + l^{\pm} + \nu$

\end{itemize}

two-body decay

\begin{itemize}
\item
$\tilde \chi^0_2 \longrightarrow \tilde l^{\pm}_{L,R} + l^{\mp}$.

\hspace{16mm} $ \hookrightarrow \tilde \chi^0_1 + l^{\pm}$

\end{itemize}

For relatively large $\tilde \chi^0_2$ mass there are two-body decays
$\tilde \chi^0_2 \longrightarrow \tilde \chi^0_1 h$,
$\tilde \chi^0_2 \longrightarrow \tilde \chi^0_1 Z$
which suppress three-body decay of $\tilde \chi^0_2$.
Direct production of $\tilde \chi^{\pm}_1 \tilde \chi^0_2$ followed
by leptonic decays of both gives three high $p_T$ isolated leptons
accompanied by missing energy due to escaping $\tilde \chi^0_1$'s and
$\nu$'s. These events do not contain jets except jets coming from
initial state radiation. Therefore the signature for 
$\tilde \chi^{\pm}_1 \tilde \chi^0_2$ pair production is 
$3l + no~jets +missing~energy$.

As mentioned above, this signature has been used in ref.~\cite{1}
for investigation of LHC gaugino discovery potential within mSUGRA model,
where gaugino masses $M(\tilde \chi^0_1)$, $M(\tilde \chi^0_2)$ are
determined mainly by a common gaugino mass $m_{1 \over 2}$ and 
$M(\tilde \chi^0_2) \approx 2.5 M(\tilde \chi^0_1)$. In our preliminary 
study we consider the general case when the relation between 
$M(\tilde \chi^{\pm}_1)$ and $M(\tilde \chi^0_1)$ is arbitrary.
We find that LHC gaugino discovery potential depends rather strongly on
the relation between $\tilde \chi^0_1$ and $\tilde \chi^0_2$ masses.

\section{Results}

Our simulations are made at the particle level with parametrized
detector responses based on a detailed detector simulation. The CMS detector
simulation program CMSJET~3.2~\cite{8} is used.
All SUSY processes with full particle spectrum, couplings,
production cross section and decays are generated with ISAJET~7.32,
ISASUSY~\cite{9}. The Standard Model backgrounds are generated 
with PYTHIA~5.7~\cite{10}. 

The following SM processes give the main contribution to the background:

\noindent
$WZ,~ZZ,~t \bar t,~Wtb,~Zb \bar b,~b \bar b$. 
In this paper we use the results of
the background simulation of ref.~\cite{1}. Namely following ref.~\cite{1}
we require 3 isolated leptons with $p^l_T > 15~GeV$ in
$|\eta^l| < 2.4~(2.5)$ for muons (electrons)
and with the same-flavour opposite-sign leptons.
As an lepton isolation criterium we require the absence of charged tracks
with $p_T > 1.5~GeV$ in a cone $R = 0.3$ around lepton. We require
also the absence of jets with $E^{jet}_T > 30~GeV$ in $|\eta^l| < 3$.
The last requirement is that the two same-flavour opposite-sign lepton
invariant mass $M_{l^+l^-} < 81~GeV$.

For such set of cuts the background cross section
$\sigma_{back} = 10^{-2}~pb$~\cite{1} that corresponds to the number
of background events $N_b = 10~(100)$ for total  luminosity
$L = 10^3~(10^4)~pb^{-1}$. See for details ref.\cite{1}. 

The results of our calculations are presented in Tables 1-4. In
estimation of the LHC gauginos discovery potential we have used
the significance determined as $S = \sqrt{N_s + N_b} - \sqrt{N_b}$
which is appropriate for the estimation of discovery potential in the case
of future experiments~\cite{11}. Here $N_s = \sigma_s \cdot L$ is the number
of signal events and $N_b = \sigma_b \cdot L$ is the number of background 
events for a given total luminosity $L$. As it follows from our results 
for given value of chargino mass $M(\tilde \chi^{\pm}_1)$ the number of
signal events depends rather strongly on the mass of the lightest 
superparticle $M(\tilde \chi^0_1)$ and for 
$M(\tilde \chi^0_1) \ge 0.6 M(\tilde \chi^{\pm}_1)$ signal is too
small to be observable. For $L = 3 \cdot 10^4pb^{-1}$ signal could
be observable for $M(\tilde \chi^{\pm}_1 ) \le 130~GeV$. 

\section{Conclusion}

In this report we have presented the results of the calculations for 
$\tilde \chi^{\pm}_1 \tilde \chi^0_2$ pair production at LHC (CMS) with their 
subsequent decays into leptons for the case of nonuniversal gaugino masses.
We have found that the visibility of signal by an excess over SM background
in $3l + no~jets + E^{miss}_T$ events depend rather strongly on the relation 
between LSP mass $\tilde \chi^0_1$ and chargino $\tilde \chi^{\pm}_1$ mass.
For total luminosity $L = 3 \cdot 10^4pb^{-1}$ signal could be 
observable for chargino mass $M(\tilde \chi^{\pm}_1) \le 130~GeV$.

\begin{center}
 {\large \bf Acknowledgments}
\end{center}

\par
We are  indebted to I.N.~Semeniouk for his help in writing the code of 
the events selections.



\begin{table}[h]
\small
    \caption{The number of events $N_{ev}$ and 
significance $S$ for  
$M(\tilde \chi^0_2)~=~104~GeV,~ M(\tilde q) = 2~TeV$,
$L~=~3 \cdot 10^4~pb^{-1}$
and for different LSP masses $M(\tilde \chi^0_1)$.}
    \label{tab.1}
\begin{center}
\begin{tabular}{|l|l|l|l|l|l|l|l|l| }
\hline
$ M(\tilde \chi^0_1)~(GeV)$ & 11 & 21 & 31 & 41 & 51 & 61 & 71 & 81 \\
\hline
$N_{ev}$                 & 181 & 152 & 331 & 242 & 179 & 121 & 21 & 4 \\
\hline
$ S $           & 4.6 & 3.9 & 7.8 & 6.0 & 4.6 & 3.2 & 0.6 & 0.12 \\
\hline
\end{tabular}
\end{center}
\end{table}

\begin{table}[h]
\small
    \caption{The number of events $N_{ev}$ and 
significance $S$ for  
$M(\tilde \chi^0_2)~=~126~GeV$, $M(\tilde q) = 2~TeV$,
$L~=~3 \cdot 10^{4}~pb^{-1}$
and for different LSP masses $M(\tilde \chi^0_1)$.}
    \label{tab:Tab.2}
\begin{center}
\begin{tabular}{|l|l|l|l|l|l|l| }
\hline
$ M(\tilde \chi^0_1)~(GeV)$ & 11 & 31 & 51 & 71 & 81 & 101  \\
\hline
$N_{ev}$                 & 25 & 19 & 124 & 59 & 35 & 3 \\
\hline
$ S $           & 0.7 & 0.5 & 3.3 & 1.6 & 1.0 & 0.1 \\
\hline
\end{tabular}
\end{center}
\end{table}

\begin{table}[h]
\small
    \caption{The number of events $N_{ev}$ and 
significance $S$ for 
$M(\tilde \chi^0_2)~=~173~GeV$, $M(\tilde q) = 2~TeV$,
$L~=~3 \cdot 10^{4}~pb^{-1}$
and for different LSP masses $M(\tilde \chi^0_1)$.}
    \label{tab:Tab.3}
\begin{center}
\begin{tabular}{|l|l|l|l|l|l|l|l|l|l|l| }
\hline
$ M(\tilde \chi^0_1)~(GeV)$ & 11 & 21 & 31 & 41 & 81 & 86 &91&96 &101&131\\
\hline
$N_{ev}$             & 3 & 10 & 7 & 13 & 19 & 88 & 74 & 89 & 43 & 25 \\
\hline
$ S $       & 0.1 & 0.3 & 0.2 & 0.4 & 0.5 & 2.4 & 2.0 & 2.4 & 1.2 & 0.7 \\
\hline
\end{tabular}
\end{center}
  \end{table}
        
\begin{table}[h]
\small
    \caption{The number of events $N_{ev}$ and 
significance $S$ for  
$M(\tilde \chi^0_2)~=~122~GeV$, $M(\tilde q) = 500~GeV$,
$L~=~3 \cdot 10^{4}~pb^{-1}$
and for different LSP masses $M(\tilde \chi^0_1)$.}
    \label{tab:Tab.4}
\begin{center}
\begin{tabular}{|l|l|l|l|l|l|l|l|l|l| }
\hline
$ M(\tilde \chi^0_1)~(GeV)$ & 9 & 19 & 29 & 39 & 51 & 58 & 79 & 88 &97\\
\hline
$N_{ev}$          & 15   & 16 & 25 & 158 & 126 & 70 & 33 & 16 & 1 \\
\hline
$ S $           & 0.4 & 0.4 & 0.7 & 4.1 & 3.3 & 1.9 & 0.9 & 0.5 & 0.03\\
\hline
\end{tabular}
\end{center}
  \end{table}
        
\end{document}